\documentclass[fleqn,usenatbib,useAMS]{mnras}

\usepackage{graphicx}	% Including figure files
\usepackage{amsmath}	% Advanced maths commands
\usepackage{amssymb}	% Extra maths symbols
\usepackage{multicol}        % Multi-column entries in tables
\usepackage{bm}		% Bold maths symbols, including upright Greek
\usepackage{pdflscape}	% Landscape pages
\usepackage{color}
\definecolor{orange}{RGB}{255, 100, 0 }
\usepackage[T1]{fontenc}
\usepackage{ae,aecompl}

\usepackage{newtxtext,newtxmath}
\title[Constraints to nodal librations by GR effects]{Constraints to the semimajor axis of outer particles with nodal librations by general relativity effects}

\author[M. Zanardi]
{M. Zanardi$^{1,2}$%
\thanks{E-mail: mzanardi@fcaglp.unlp.edu.ar},
G. C. de El\'ia$^{1,2}$, 
A. Dugaro$^{1,2}$, 
and C. F. Coronel$^{1,2}$
\\
$^{1}$Instituto de Astrof\'{\i}sica de La Plata, CCT La Plata-CONICET-UNLP, Paseo del Bosque S/N (1900), La Plata, Argentina\\ 
$^{2}$Facultad de Ciencias Astron\'omicas y Geof\'{\i}sicas, Universidad Nacional de La Plata, Paseo del Bosque S/N (1900), La Plata, Argentina
}

\date{Accepted XXX. Received YYY; in original form ZZZ}

\pubyear{2023}

\begin{document}
\label{firstpage}
\pagerange{\pageref{firstpage}--\pageref{lastpage}}
\maketitle

\begin{abstract}
%This is a guide for preparing papers for \textit{Monthly Notices of the Royal Astronomical Society} using the \verb'mnras' \LaTeX\ package.
%It provides instructions for using the additional features in the document class.
%This is not a general guide on how to use \LaTeX, and nor does it replace the journal's instructions to authors.
%See \texttt{mnras\_template.tex} for a simple template.
We study nodal librations of outer particles in the framework of the elliptical restricted three-body problem including general relativity (GR) effects. From an analytical treatment based on secular interactions up to quadrupole level, we derive equations that define the nodal libration region of an outer test particle, which depends on the physical and orbital parameters of the bodies of the system. From this, we analyze how the GR constrains the semimajor axis of outer test particles that experience nodal librations under the effects of an inner planet around a single stellar component. Such an upper limit of the semimajor axis, which is called $a_{2,\text{lim}}$, depends on the mass of the star $m_{\text{s}}$, the mass $m_1$, the semimajor axis $a_1$, and the eccentricity $e_1$ of the inner planet, and the eccentricity $e_2$ of the outer test particle. On the one hand, our results show that the greater $m_1$, $a_1$, and $e_2$ and the smaller $m_{\text{s}}$, the greater the value of $a_{2,\text{lim}}$. On the other hand, for fixed $m_{\text{s}}$, $m_1$, $a_1$, and $e_2$,  $a_{2,\text{lim}}$ does not strongly depend on $e_1$, except for large values of such an orbital parameter. We remark that N-body experiments of particular scenarios that include GR show results consistent with the analytical criteria derived in the present research. Moreover, the study of hypothetical small body populations of real systems composed of a single star and an inner planetary-mass companion show that the GR effects can play a very important role in their global dynamics. 
\end{abstract}


%%%%%%%%%%%%%%%%%%%%%%%%%%%%%%%%%%%%%%%%%%%%%%%%%%

%%%%%%%%%%%%%%%%% BODY OF PAPER %%%%%%%%%%%%%%%%%%

% The MNRAS class isn't designed to include a table of contents, but for this document one is useful.
% I therefore have to do some kludging to make it work without masses of blank space.
%\begingroup
%\let\clearpage\relax
%\tableofcontents
%\endgroup
%\newpage

\section{Introduction}
The wide diversity of planetary systems found in the vicinity of the Sun shows different dynamical properties depending on the physical and orbital parameters of the bodies that compose them. In particular, the global dynamical structure of potential outer small body populations associated with single-planet systems orbiting an only stellar component is complex. In fact, if the planet is located at inner regions of the system, the dynamical features of those potential outer reservoirs will be strongly governed by general relativity (GR) effects. Recently, our investigations were aimed at addressing this issue, studying the dynamics of outer particles that evolve under the effects of an inner and eccentric perturber in the framework of the elliptical restricted three-body problem with and without GR effects. 

The pioneer study aimed at analyzing the secular orbital evolution of a planet in a binary star system was developed by \cite{Ziglin1975}. In such a research, the author worked up to the quadrupole level of the secular theory, showing that, if the planet is assumed to be a test particle, its orbital inclination experiences a coupling with the ascending node longitude, which can librate or circulate. In particular, \citet{Ziglin1975} showed that the nodal libration regime defines an orbit-flipping resonance, where the test particle's inclination symmetrically oscillates between prograde and retrograde values around 90$^{\circ}$, where the extremes of the ascending node longitude are obtained. Furthermore, the author derived an analytical expression that allows to obtain the extreme values of the orbital inclination of the outer particle on a nodal libration trajectory, which only depend on the eccentricity of the inner perturber. More recently, this orbit-flipping resonance has been study by many authors in the context of the elliptical restricted three-body problem from the development of analytical criteria and numerical experiments \citep[e.g.][]{Verrier2009, Farago2010, Li2014, Naoz2017, Zanardi2017, Vinson2018}.  

\cite{Zanardi2018} carried out for the first time a detailed study about how the GR effects modify the dynamical behavior of an outer test particle in the framework of the elliptical restricted three-body problem. In such a research, the authors worked on the basis of a secular Hamiltonian up to quadrupole level and derived an integral of motion of the system with GR, which is conserved over the individual evolutionary trajectory of the outer particle. From this, \citet{Zanardi2018} found an analytical expression that defines the nodal libration region, which depends on the mass of the central star, the mass, the semimajor axis, and the eccentricity of the inner perturber, as well as the semimajor axis and the eccentricity of the test particle. From this analysis, the authors showed that the GR reduces the range of prograde inclinations of the nodal libration region with respect to that obtained in absence of GR. Moreover, they found that the extreme values of the ascending node longitude are obtained for retrograde inclinations, which leads to two different nodal libration regimes for the outer test particle. On the one hand, nodal librations associated with orbital flips, where the inclination asymmetrically oscillates between prograde and retrograde values around 90$^{\circ}$. On the other hand, nodal librations correlated with purely retrograde orbits. According to this work, when the perturber is located at inner regions of the system, the GR must be included in the models of evolution since its effect governs the outer global dynamical structure.

Recently, \cite{Lepp_Martin2022} studied how the GR affects the dynamics of circumbinary planets in wide orbits around a close binary. In the context of the elliptical restricted three-body problem, the authors worked with analytical criteria derived by \citet{Zanardi2018} and found that there is a critical radius beyond which an outer test particle follows nodal circulation trajectories for any value of its initial inclination. Moreover, \citet{Lepp_Martin2022} showed that this radial limit depends on the physical and orbital properties of the binary. It is important to mention that the authors restricted their study to an inner binary considering stellar masses for both components and a circular orbit for the circumbinary test particle. 

From the catalog of 2023 July 4, about 70 \% of confirmed planetary systems are composed of an only planet with a minimum mass between about 0.1 M$_{\oplus}$ and 13 M$_{\text{Jup}}$, which orbits a single stellar component\footnote{https://exoplanetarchive.ipac.caltech.edu}. In that sample, the
mass of the central object ranges from the substellar limit to a few solar masses, showing a strong concentration around a solar mass. Moreover, most of those planets have a semimajor axis less than 1 au, with a peak in the distribution around 0.05 au. Finally, the orbital eccentricities listed in the sample are preferentially low, although they are distributed over a wide range of values with a maximum of about 0.93.
According to the physical and orbital properties of the bodies that compose those systems of the observed sample, the GR effects will play a key role in the evolution of the planetary companion, which will have important implications in the dynamics of potential outer small body reservoirs \citep{Zanardi2018}. Motivated by this, the main goal of the present work is to analyze how the GR constrains the range of semimajor axes of the nodal libration region of an outer test particle in the elliptical restricted three-body problem. In this line of research, we are interested at studying the sensitivity of the results to the mass of the central star and to the mass, semimajor axis, and eccentricity of the inner perturber for the range of values of the observed sample. Moreover, we want to analyze the dependency of our results on the eccentricity adopted for the outer test particle, which is very important for understanding the structure and evolution of hypothetical outer reservoirs of small bodies associated with real exoplanetary systems.

The present investigation is organized as follows. In Sect. 2, we introduce the analytical treatment from which we develop the present study. The results obtained from the use of the analytical prescriptions are presented in Sect. 3. The numerical results concerning the dynamical properties of the outer test particle in the framework of the elliptical restricted three-body problem with application to real systems are shown in Sect.~4. Finally, the discussions and conclusions of the present research are exposed in Sect.~5. 

\section{Method}
In this section, we present the analytical method used to study the role of the GR in the dynamics of test particles that evolve under the influence of an inner and eccentric massive perturber in the elliptical restricted three-body problem. In particular, we describe the analytical treatment based on the Hamiltonian of the outer test particle expanded up to quadrupole level of the secular approximation with the inclusion of the GR effects proposed by \citet{Zanardi2018}. 

Following to \cite{Ziglin1975} and \cite{Naoz2017}, the secular Hamiltonian up to quadrupole level of the outer test particle in the framework of the elliptical restricted three-body problem is given by 
\begin{eqnarray}
  f_{\text{quad}} = \frac{(2+3e^{2}_{1})(3\cos^{2}i_{2}-1)+15e^{2}_{1}(1-\cos^{2}i_{2})\cos2\Omega_2}{(1-e^{2}_{2})^{3/2}}
  \label{eq:fquad}
\end{eqnarray}       
where $e_1$ is the inner perturber's eccentricity, and $e_2$, $i_2$, and $\Omega_2$ are the eccentricity, the inclination, and the ascending node longitude of the outer test particle, respectively. It is important to remark that the inclination $i_2$ is referenced to the invariant plane of the system, while the ascending node longitude $\Omega_2$ is measured respect to the pericentre of the inner perturber. 

The GR generates a precession of the inner orbit with a nomimal rate given by
\begin{eqnarray}
    \dot{\omega}_{1,\textrm{GR}} = \frac{3 k^3 (m_\textrm{s} + m_1)^{3/2}}{c^2 a^{5/2}_1 (1-e^2_1) }
    \label{eq:Einstein}
\end{eqnarray}
\citep{Einstein1916} where $\omega_1$ is the argument of pericentre of the inner planet, $k^2$ the gravitational constant, $c$ the speed of light, $m_{\text{s}}$ and $m_1$ the mass of the star and the inner perturber, respectively, and $a_1$ and $a_2$ the semimajor axis of the inner perturber and the test particle, respectively. This effect translates to a precession of the ascending node longitude of the outer test particle with a rate given by $\dot{\Omega}_{2,\textrm{GR}} = - \dot{\omega}_{1,\textrm{GR}}$, where $\dot{\omega}_{1,\textrm{GR}}$ refers to Eq.~\ref{eq:Einstein} \citep{Naoz2017}. From this, \citet{Zanardi2018} included the GR effects in the analytical treatment of the elliptical restricted three-body problem for the case of an outer test particle. The authors worked in the rotating frame of the inner planet's orbit and found that the GR leads to an integral of motion, which adopts the expression
\begin{eqnarray}
f = f_{\text{quad}} + 48 \frac{k^2}{c^2}\frac{(m_{\textrm{s}}+m_{1})^3}{m_{\textrm{s}}m_{1}} \frac{(1-e^{2}_{2})^{1/2}}{(1-e^{2}_{1})} \frac{a^{7/2}_{2}}{a^{9/2}_{1}} \cos{i_{2}}, 
\label{eq:f-energia}
\end{eqnarray}  
where $f_{\textrm{quad}}$ is given by Eq.~\ref{eq:fquad}. 

\begin{figure*}
  \centering
  \includegraphics[angle=0, width=0.48\textwidth]{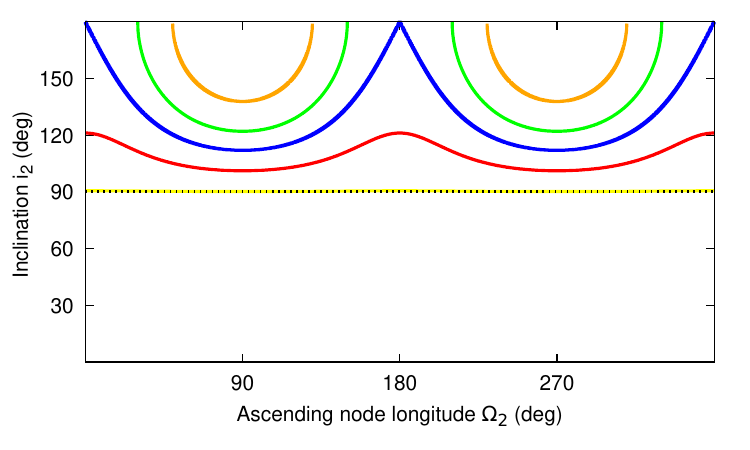}
  \includegraphics[angle=0, width=0.48\textwidth]{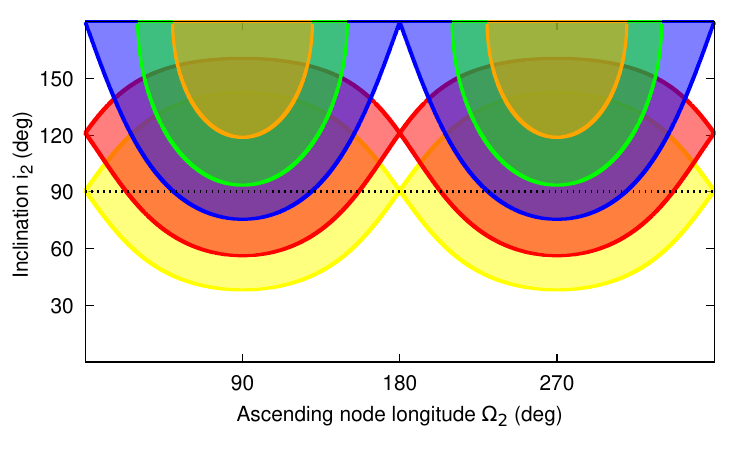}
   \caption{
   Left panel: Values of $i^{*}_2$ as a function of $\Omega_2$ obtained from Eq.~\ref{eq:iestre}. Right panel: Extreme inclinations (curves) and nodal libration regions (color shaded zones) of an outer test particle derived from Eqs.~\ref{eq:cuadratica} and \ref{eq:cuadratica_nueva}. In all cases, the system assumes $m_{\text{s}} =$ 1 M$_{\odot}$, $m_1 =$ 1 M$_{\text{Jup}}$, $a_1 =$ 1 au, and $e_1 = e_2 =$ 0.5. The yellow, red, blue, green, and orange colors are associated with a value of $a_2$ of 5 au, 15 au, 18.105 au, 20 au, and 22 au, respectively.
   }
\label{fig:fig_nueva}
\end{figure*}

\cite{Zanardi2018} described the necessary requirements for the production of nodal librations of an outer test particle, finding significant differences with respect to that obtained in absence of GR. In fact, when the GR is not included in the model, the nodal libration trajectories of an outer test particle are correlated with symmetric orbital flips around $i_2 = 90^\circ$, where the extreme values of $\Omega_2$ always occur for any scenario of work. However, when the GR is taken into account, the extreme values of $\Omega_2$ occur at a value of $i_2 = i^{*}_2$, which depends on the physical and orbital properties of the bodies of the system. In this case, the nodal libration trajectories are asymmetric around $i^{*}_2$, which is given by
\begin{eqnarray}
i^{*}_2 = \arccos{ \left ( A \frac{(1-e^{2}_{2})^{2}} {(1-e^{2}_{1}) (2 + 3 e^{2}_{1} - 5 e^{2}_{1} \cos 2 \Omega_2)} \frac{a^{7/2}_{2}}{a^{9/2}_{1}} \right)}, 
\label{eq:iestre}
\end{eqnarray}  
being $A$ a constant whose expression is 
\begin{eqnarray}
 A = -\frac{8 k^2 (m_{\textrm{s}} + m_{1})^3}{c^2 m_{\text{s}} m_1}.  
 \label{eq:parametro_A}
\end{eqnarray}
\noindent{For a system characterized by the parameters} $m_{\textrm{s}}$, $m_1$, $a_1$, $e_1$, $a_2$, and $e_2$, the existence of $i^{*}_2$ for a range of values of $\Omega_2$ guarantees that the outer test particle can experience nodal librations for an appropriate value of $i_2$. In the context of our research, it is important to analyze the dependence of $i^{*}_2$ on the outer test particle's semimajor axis $a_2$. The left panel of Fig.~\ref{fig:fig_nueva} illustrates five examples of $i^{*}_2$ as a function of $\Omega_2$ for $m_{\textrm{s}} =$ 1 M$_{\odot}$, $m_1 =$ 1 M$_{\textrm{Jup}}$, $a_1 =$ 1 au, and $e_1 = e_2 =$ 0.5. In particular, the yellow, red, blue, green, and orange curves represent the values of $i^{*}_2$ for $a_2$ of 5 au, 15 au, 18.105 au, 20 au, and 22 au, respectively. From values of $a_2$ comparable to $a_1$ up to a given particular value $a_{2,\text{p}}$, $i^{*}_2$ is defined for any $\Omega_2$ between 0$^{\circ}$ and 360$^{\circ}$. In our example, this behavior is observed for $a_2 =$ 5 au, 15 au, and 18.105 au, the latter being the $a_{2,\text{p}}$ for which $i^{*}_2 =$ 180$^{\circ}$ at $\Omega_2 = 0^{\circ}$. As Eq.~\ref{eq:iestre} shows, $i^{*}_2$ is an increasing function of $a_2$ at a fixed $\Omega_2$, for which $i^{*}_2$ has no solution at $\Omega_2 = 0^{\circ}$ for $a_2 > a_{2,\text{p}}$. In our example, this occurs for $a_2 =$ 20 au and 22 au, where $i^{*}_2 = 180^{\circ}$ for $\Omega_2 = 30^{\circ}$ and 50$^{\circ}$, respectively. From our analysis, we derive two important conclusions. First, if $i^{*}_2$ has no solution at $\Omega_2 = 0^{\circ}$, the generation of nodal librations requires the existence of a value of $\Omega_2 = {\hat \Omega}_2$ for which $i^{*}_2 = 180^{\circ}$. Second, for $a_2 > a_{2,\text{p}}$, the greater the $a_2$, the smaller the range of values of $\Omega_2$ around $\pm 90^{\circ}$ for which $i^{*}_2$ is defined.

In this framework, it is necessary to develop a procedure that allows us to obtain the nodal libration region for those scenarios where $i^{*}_2$ exists at $\Omega_2 = 0^{\circ}$ and for those in which it does not. To do this, it is important to remember that the integral of motion $f$ given by Eq.~\ref{eq:f-energia} is conserved on the evolutionary trajectory of any outer test particle in the elliptical restricted three-body problem based on a secular and quadrupole Hamiltonian with GR. Making use of $f$, \citet{Zanardi2018} determined the extreme inclinations $i^{\textrm{e}}_2$ that lead to nodal librations of an outer test particle, which are associated with $\Omega_2 = \pm 90^{\circ}$. On the one hand, if $i^{*}_2$ exists for $\Omega_2 =$ 0$^\circ$, $f(\Omega_2 = \pm 90^{\circ}, i_2 = i^{\textrm{e}}_2) = f(\Omega_2 = 0^{\circ}, i_2 = i^{*}_2)$ is proposed, from which the extreme values $i^{\textrm{e}}_2$ are obtained by solving the following quadratic equation
\begin{eqnarray}
  \alpha \cos^{2} i^{\text{e}}_{2} + \beta \cos i^{\text{e}}_{2} + \gamma = 0,
  \label{eq:cuadratica}
\end{eqnarray}
\noindent{where} the coefficients are given by 
\begin{eqnarray}
 \label{eq:alfa}
 \alpha &=& 1 + 4 e^{2}_{1}, \\ 
 \label{eq:beta}
 \beta &=& -A
 \frac{(1-e^{2}_{2})^{2}}{(1-e^{2}_{1})} \frac{a^{7/2}_{2}}{a^{9/2}_{1}}, \\ 
\label{eq:gama}
 \gamma &=& \frac{\beta^{2}}{4(1-e^{2}_{1}) } - 5e^{2}_{1}.
\end{eqnarray}
\noindent{On} the other hand, if $i^{*}_2$ does not exist for $\Omega_2 = 0^{\circ}$, we propose $f(\Omega_2 = 90^{\circ}, i_2 = i^{e}_2) = f(\Omega_2 = \hat{\Omega}_{2}, i_2 = i^{*}_2 = 180^{\circ})$ \footnote{It is worth noting that if the function $f$ is evaluated at $i_2 = 180^{\circ}$, its value does not depend on $\Omega_2$.}. From this, the extreme values of the inclination that lead to nodal libration trajectories are obtained from the following quadratic function 
\begin{eqnarray}
  \alpha \cos^{2} i^{\text{e}}_{2} + \beta \cos i^{\text{e}}_{2} + \gamma' = 0,
  \label{eq:cuadratica_nueva}
\end{eqnarray}
\noindent{where} $\alpha$ and $\beta$ are given by Eqs.~\ref{eq:alfa} and \ref{eq:beta}, respectively, and the $\gamma'$ coefficient is given by
\begin{eqnarray}
    \gamma' = \beta - \alpha.
    \label{eq:gamma2}
\end{eqnarray}

\noindent{In this case,} it is possible to verify that the maximum extreme inclination $i_{2,\textrm{max}}^{\textrm{e}}$ is equal to 180$^{\circ}$ while the minimum extreme inclination $i_{2,\textrm{min}}^{\textrm{e}}$ is written as 
\begin{eqnarray}
i_{2,\textrm{min}}^{\textrm{e}} = \textrm{arccos} \left(1 - \frac{\beta}{\alpha} \right).
\label{eq:i2min_nueva_cuadratica}
\end{eqnarray}

\noindent{The right} panel of Fig.~\ref{fig:fig_nueva} illustrates the nodal libration region for $a_2$ of 5 au (yellow), 15 au (red), 18.105 au (blue), 20 au (green), and 22 au (orange) in a ($\Omega_2$, $i_2$) plane. On the one hand, the extreme inclinations are calculated from Eq.~\ref{eq:cuadratica} for $a_2 =$ 5 au and 15 au, while Eq.~\ref{eq:cuadratica_nueva} is used for $a_2 =$ 20 au and $a_2 =$ 22 au. On the other hand, given that $i^{*}_2 = 180^{\circ}$ at $\Omega_2 = 0^{\circ}$ for $a_2 = a_{2,\text{p}} =$ 18.105 au, the extreme inclinations can be calculated either by Eq.~\ref{eq:cuadratica} or by  \ref{eq:cuadratica_nueva}. From this, for $a_2 > a_{2,\text{p}}$, the use of  Eq.~\ref{eq:cuadratica_nueva} shows that the greater the outer particle's semimajor axis $a_2$, the higher the minimum extreme inclination $i^{\text{e}}_{2,\text{min}}$ and the smaller the range of values of $\Omega_2$ around $\pm 90^{\circ}$ that lead to nodal librations. 

We remark that the GR effects produce a nodal libration region for an outer test particle, which strongly depends on the orbital elements and physical properties of the bodies that compose the system.

\section{Analytical results}
Here, we present the results obtained from analytical criteria considering GR effects on the dynamical evolution of an outer particle in single-planet systems with an only stellar component. Based on a secular theory up to the quadrupole level of the approximation, we are particularly interested in providing constraints on the semimajor axis of the outer test particle below which nodal librations can be experienced. Moreover, we analyze the sensitivity of our results to the physical and orbital parameters involved in the systems under study within the range of values derived from the observed sample.

\subsection{Semimajor axis upper limit for nodal librations}

Such as we mentioned in the previous section, for a set of parameters $m_{\textrm{s}}$, $m_1$, $a_1$, $e_1$, $a_2$, and $e_2$, 
the existence of $i^{*}_2$ for a given range of values of $\Omega_2$ guarantees that the outer particle can experience nodal librations for appropriate values of $i_2$. If $i^{*}_2$ exists for $\Omega_2 = 0^{\circ}$, Eq.~\ref{eq:iestre} shows that such a value is an increasing function of $a_2$ and the extreme inclinations that lead to nodal librations are calculated from Eq.~\ref{eq:cuadratica}. From this, there is a particular value of $a_2$ called $a_{2,\text{p}}$ for which $i^{*}_2 = 180^{\circ}$ at $\Omega_2 = 0^{\circ}$. Finally, if we keep increasing the outer particle's semimajor axis $a_2$ above $a_{2,\text{p}}$, $i^{*}_2$ does not exist for $\Omega_2 = 0^{\circ}$, the range of values of $\Omega_2$ around $\pm 90^{\circ}$ where $i^{*}_2$ has solution decreases, and the minimum extreme inclination of the nodal libration region calculated from Eq.~\ref{eq:i2min_nueva_cuadratica} increases. From this description, there exists a limiting case for which $i^{*}_2$ is only defined at $\Omega_2 = \pm 90^{\circ}$ with a value of 180$^{\circ}$, and the minimum and maximum extreme inclinations of the nodal libration region are also both equal to 180$^{\circ}$. By adopting the values ($\Omega_2 = \pm 90^{\circ}$, $i^{*}_2 = 180^{\circ}$) at Eq.~\ref{eq:iestre} or $i^{\textrm{e}}_{2,\textrm{min}} = 180^{\circ}$ at Eq.~\ref{eq:i2min_nueva_cuadratica}, we derive an upper limit of $a_2$ beyond which the outer test particle can not experience nodal librations. This value, which is called $a_{2,\textrm{lim}}$, is given by     
\begin{eqnarray}
a_{2,\text{lim}} = \left( \frac{-2 (1-e^{2}_{1}) (1 + 4 e^{2}_{1})} {A (1 - e^{2}_{2})^{2}} a^{9/2}_{1}  \right)^{2/7}.
 \label{eq:a2}
\end{eqnarray}

\noindent{From this expression, it is clear that the} value of $a_{2,\text{lim}}$ explicitly depends on the mass of the star $m_{\text{s}}$, the mass $m_1$, the semimajor axis $a_1$, and the eccentricity $e_1$ of the inner perturber, and the eccentricity $e_2$ of the outer test particle. We remark that \cite{Lepp_Martin2022} recently derived a particular case of Eq.~\ref{eq:a2} associated with $e_2 = 0$ in the framework of an outer test particle orbiting a stellar binary. 

In the context of our research, we are interested in studying the sensitivity of $a_{2,\text{lim}}$ to the physical and orbital parameters of the bodies that compose the system of work. Thus, a detailed study of these dependencies is developed in the following subsections.

\subsection{Nodal librations: range of the inclinations of the outer test particle}

\label{sec:lib_node}
In this section, we are interested at analyzing the range of inclinations for which the outer test particle can follow nodal libration trajectories as a function of its semimajor axis $a_2$. To do this, we consider a system composed of a solar-mass star, a Jupiter-mass planet with a semimajor axis $a_1 =$ 1~au and an eccentricity $e_1 =$ 0.5, and an outer test particle with different values of its semimajor axis $a_2$ and eccentricity $e_2$. As we described in Sect. 2, the extreme inclinations that lead to nodal librations are obtained by solving Eqs.~\ref{eq:cuadratica} or \ref{eq:cuadratica_nueva} depending on the existence of $i^{*}_2$ at $\Omega_2 = 0^{\circ}$.
\begin{figure}
  \centering
  \includegraphics[angle=0, width=0.48 \textwidth]{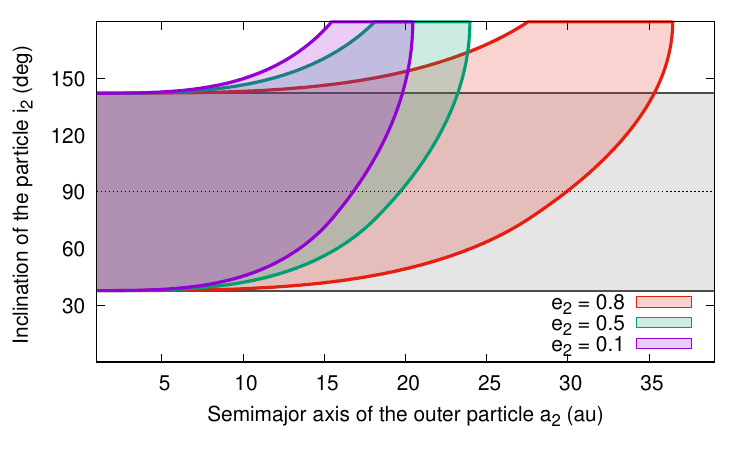}
   \caption{The violet, green, and red curves illustrate the extreme inclinations of the nodal libration region of an outer test particle as a function of its semimajor axis $a_2$ for values of $e_2$ of 0.1, 0.5, and 0.8, respectively, when GR effects are included. The horizontal gray lines represent the extreme inclinations of the nodal libration region in absence of GR. The color shaded regions illustrate the values of $i_2$ at $\Omega_2 = \pm 90^{\circ}$ and $a_2$ that lead to nodal libration trajectories of the outer test particle in the different scenarios of work. In all cases, the massive bodies of the system are represented by a solar-mass star and a Jupiter-mass planet with $a_1 =$ 1~au and $e_1 =$ 0.5.}
\label{fig:fig.a2_iext.eps}
\end{figure}

Figure~\ref{fig:fig.a2_iext.eps} shows the orbital inclinations of an outer test particle associated with nodal libration trajectories as a function of $a_2$ for different values of $e_2$ with and without GR effects. On the one hand, the gray shaded region represents the values of the inclination $i_2$ at $\Omega_2 = \pm 90^{\circ}$ for which an outer test particle can experience nodal librations in absence of GR. Note that this region does not depend on either $a_2$ or $e_2$ since it is only a function of $e_1$. In fact, according to \citet{Ziglin1975}, 

the extreme values of $i_2$ associated with nodal librations are obtained by
\begin{eqnarray}
i^{\text{e}}_{2,\text{No-GR}} =  \arccos \left\lbrace {\pm \sqrt {\frac{5 e^{2}_{1}}{1+4e^{2}_{1}}}}\right\rbrace,
\label{eq:incli_ziglin}
\end{eqnarray}
which are illustrated by the horizontal gray lines in Fig.~\ref{fig:fig.a2_iext.eps}. On the other hand, the color curves of such a figure represent the extreme values of $i_2$ with GR obtained from Eqs.~\ref{eq:cuadratica} and \ref{eq:cuadratica_nueva}, while the corresponding color shaded regions show the $i_2$ values at $\Omega_2 = \pm 90^{\circ}$ that lead to nodal libration trajectories for each pair of parameters $e_2$ and $a_2$.   

Figure~\ref{fig:fig.a2_iext.eps} shows that the GR effects significantly modifies the orbital inclinations that leads to an outer test particle to follow trajectories of nodal libration respect to that obtained in absence of GR. As Eq.~\ref{eq:incli_ziglin} shows, the extreme inclinations that lead to nodal librations without GR effects only depend on $e_1$. When GR effects are included, the values of $i^{\text{e}}_2$ associated with the nodal libration region depends on $m_{\text{s}}$, $m_1$, $a_1$, $e_1$, $a_2$ and $e_2$. This can be observed from the explicit dependence of the coefficients $\alpha$, $\beta$, $\gamma$, and $\gamma'$ of Eqs.~\ref{eq:cuadratica} and \ref{eq:cuadratica_nueva} on those physical and orbital parameters. From the particular scenario of study illustrated in Fig.~\ref{fig:fig.a2_iext.eps}, it is observed that the extreme inclinations of the nodal libration region obtained with GR are comparable to those derived without GR for small values of $a_2$. However, an increase in $a_2$ produces significant changes in the extreme inclination values that lead to nodal librations with GR compared to those derived without GR. In this sense, Fig.~\ref{fig:fig.a2_iext.eps} shows that the minimum and maximum extreme inclinations of the nodal libration region with GR become greater with an increment of $a_2$ until each of them reaches a value of 180$^{\circ}$. 

From this analysis and such as we mentioned in Sect. 3.1, Fig.~\ref{fig:fig.a2_iext.eps} shows that the GR effects determine an upper limit of $a_2$ beyond which nodal librations of the outer test particle are not possible. According to this, such an upper limit of $a_2$ corresponds to $i^{\text{e}}_{2,\text{min}} = i^{\text{e}}_{2,\text{max}}= 180^{\circ}$ and its value is that given by Eq.~\ref{eq:a2}. The particular analysis showed in Fig.~\ref{fig:fig.a2_iext.eps} indicates that such a limit value of $a_{2}$ is strongly dependent on $e_2$. In fact, the greater the $e_2$ value, the larger the upper limit of $a_{2}$ below which nodal librations are possible. A simple analysis of Eq.\ref{eq:a2} makes this dependency clear.
 \label{eq:rango}
 \label{eq:rango_2}

\subsection{Sensitivity to the mass of the inner perturber}

Here, we study the values of the semimajor axis of an outer test particle associated with nodal libration trajectories as a function of the inner perturber's mass $m_1$. From the analysis of $A$ parameter given by Eq.~\ref{eq:parametro_A}, Eq.~\ref{eq:a2} indicates that the upper limit of the semimajor axis $a_{2,\text{lim}}$ is an increasing function of $m_1$.  

\begin{figure}
  \centering
  \includegraphics[angle=0, width=0.5
  \textwidth]{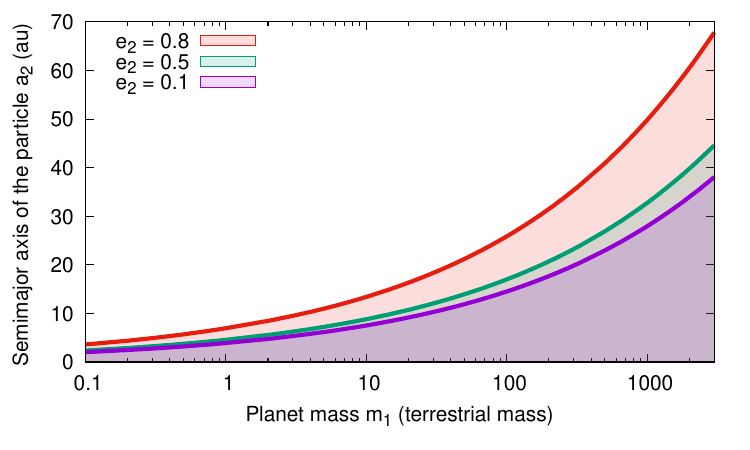}
  \caption{Values of $a_{2,\text{lim}}$ as a function of the inner perturber's mass $m_1$ for $e_2 =$ 0.1 (violet curve), 0.5 (green curve), and 0.8 (red curve). The color shaded regions illustrate the ($m_1$, $a_2$) pairs that can lead to nodal librations of the outer test particle in the different scenarios of work. In all cases, the system assumes $m_{\text{s}} =$ 1 M$_{\odot}$, $a_1 =$ 1 au, and $e_1 =$ 0.5.}
\label{fig:fig.a2_m1}
\end{figure}

To study this, we consider a system composed of a central star of 1~M$_{\odot}$, an inner perturber of mass $m_1$ with a semimajor axis $a_1 =$ 1 au and an eccentricity $e_1 =$ 0.5, and an outer test particle with different values of the eccentricity $e_2$. Figure~\ref{fig:fig.a2_m1} illustrates the nodal libration region of an outer test particle as a function of $a_2$ and $m_1$ in our system under consideration, when the GR effects are included. On the one hand, the violet, green, and red curves represent the upper limit of the semimajor axis $a_{2,\text{lim}}$ as a function of $m_1$ for $e_2 =$ 0.1, 0.5, and 0.8, respectively, which is calculated by Eq.~\ref{eq:a2}. On the other hand, the corresponding color shaded regions show the pairs ($m_1$, $a_2$) associated with nodal librations of the outer test particle. As the reader can see in such a figure, the more massive the inner planet, the greater the upper limit $a_{2,\text{lim}}$ below which the test particle can experience nodal librations. 

This result has very important implications over the entire range of planetary masses. In fact, systems hosting a single inner terrestrial-like planet are of special interest because the GR suppresses the nodal librations of any outer test particle located only a few astronomical units away from such a planet. For our particular scenario of study, an outer particle with $e_2 =$ 0.1, 0.5, and 0.8 that evolves under the influence of a super-Earth of 5~M$_{\oplus}$ can not follow nodal libration trajectories for values of $a_2$ greater than 6.2 au, 7.2 au, and 11 au, respectively. Thus, this kind of systems could only host Kuiper belt analogue reservoirs with nodal circulations\footnote{In the present paper, we call Kuiper belt analogue to those small body populations with a location similar to the mentioned reservoir of our Solar System.}. From Fig.~\ref{fig:fig.a2_m1}, the existence of a Kuiper belt analogue with nodal librations beyond 30 au is only associated with an inner perturber more massive than 4.2 M$_\text{Jup}$, 2.3 M$_\text{Jup}$, and 0.53 M$_\text{Jup}$ for $e_2$ of 0.1, 0.5, and 0.8, respectively. From this, it is worth noting that distant small body populations slightly excited in eccentricity around solar-mass stars could only experience nodal librations under the effects of an inner massive super-Jupiter in our scenario of work.

\subsection{Sensitivity to the stellar mass}

Now, we analyze the sensitivity of the nodal libration region of an outer test particle to the mass of the central star $m_{\text{s}}$. As we mentioned in the previous section, from a simple analysis of $A$ parameter given by Eq.~\ref{eq:parametro_A}, Eq.~\ref{eq:a2} shows that $a_{2,\text{lim}}$ decreases with an increase in the value of $m_{\text{s}}$.

\begin{figure}
  \centering
  \includegraphics[angle=0,width=0.5\textwidth]{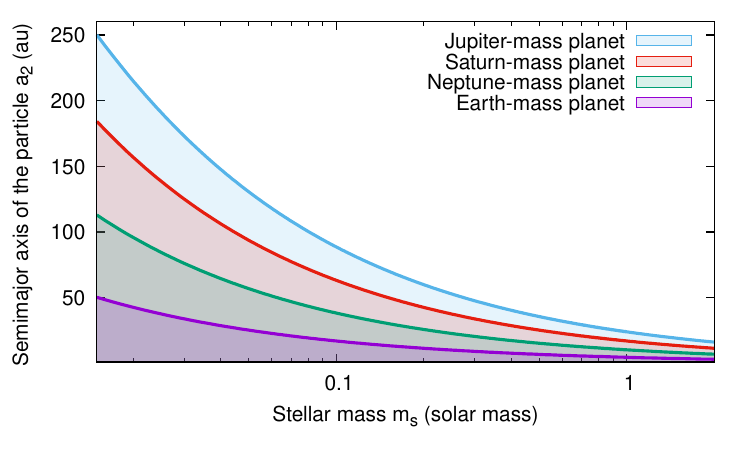}
  \caption{
Values of $a_{2,\text{lim}}$ as a function of the stellar mass $m_{\text{s}}$ for an Earth- (violet curve), a Neptune- (green curve), a Saturn- (red curve), and a Jupiter-mass planet (sky blue curve). The color shaded regions illustrate the ($m_{\text{s}}$, $a_2$) pairs that can lead to nodal librations
of the outer test particle in the different scenarios of work. In all cases, the system assumes $a_1 =$ 1 au, and $e_1 = e_2 =$ 0.5.
  }
\label{fig:fig.ms_a2}
\end{figure}

To analyze this dependency, we define a scenario of work by considering an inner perturber with different values of the mass $m_1$, a semimajor axis $a_1 =$ 1 au, and an eccentricity $e_1 =$ 0.5, and an outer test particle with an eccentricity $e_2 =$ 0.5. 
The violet, green, red, and sky blue curves showed in Fig.~\ref{fig:fig.ms_a2} illustrate the upper limit of the semimajor axis $a_{2,\text{lim}}$ as a function of the $m_{\text{s}}$ for an Earth-, Neptune-, Saturn-, and Jupiter-mass planet, respectively. Moreover, the color shaded regions represent the ($m_\textrm{s}$, $a_2$) pairs that can produce nodal librations of the outer test particle of our scenario of study for the different masses associated with the inner perturber. Our results show that, for a fixed value of $m_1$, the more massive the central star, the smaller the upper limit $a_{2,\textrm{lim}}$ below which nodal librations of the outer test particle are possible. In fact, for the particular case of an inner Jupiter-mass planet, the value of $a_{2,\textrm{lim}}$ is equal to 87.5 au and 23.65 au for $m_{\text{s}} =$ 0.1 M$_{\odot}$ and 1 M$_{\odot}$, respectively.

It is interesting to analyze the results of Fig.~\ref{fig:fig.ms_a2} concerning the dynamical properties of potential Kuiper belt-like reservoirs beyond 30 au in the scenario under consideration. On the one hand, Kuiper belt analogues that evolve under the effects of an inner Earth-mass planet can only experience nodal circulations for any value of $m_{\text{s}}$ above the substellar limit. From the violet curve illustrated in Fig.~\ref{fig:fig.ms_a2}, it is important to remark that nodal librations of such potential reservoirs can only be associated with a brown dwarf less massive than 0.038 M$_{\odot}$. On the other hand, an inner Neptune-, Saturn-, and Jupiter-mass planet only allows the existence of a Kuiper belt analogue reservoir with nodal librations for a central star less massive than 0.16 M$_{\odot}$, 0.36 M$_{\odot}$, and 0.66 M$_{\odot}$, respectively.

\subsection{Sensitivity to the eccentricity and semimajor axis of the inner perturber}

Here, we analyze the possible values of the semimajor axis $a_2$ that lead to nodal libration trajectories of an outer test particle as a function of the semimajor axis $a_1$ and eccentricity $e_1$ of the inner perturber.

\begin{figure}
  \centering
  \includegraphics[angle=0, width=0.5\textwidth]{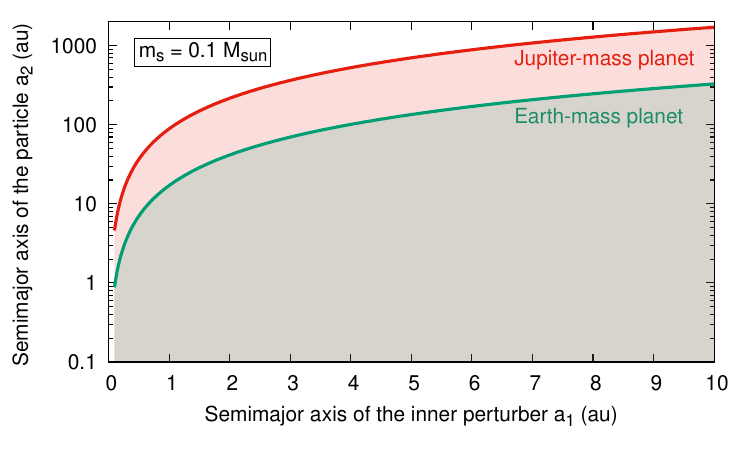} \\
\includegraphics[angle=0, width=0.5\textwidth]{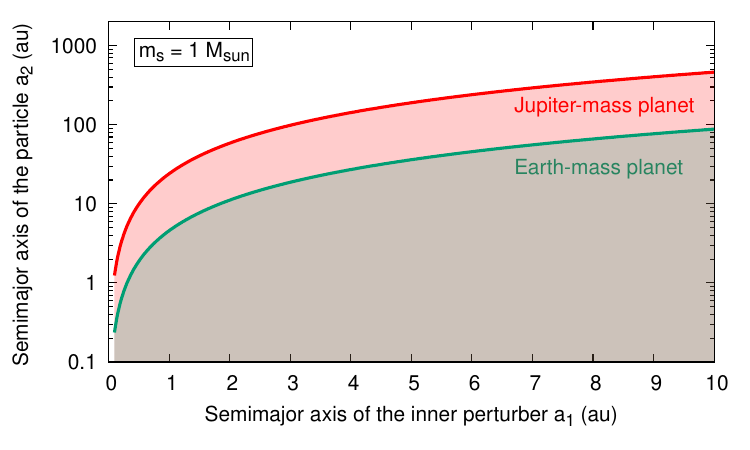}
  \caption{
Values of $a_{2,\text{lim}}$ as a function of the inner perturber’s semimajor axis $a_1$ for an Earth- (green curve), and a Jupiter-mass planet (red curve), by assuming a central star of 0.1 M$_{\odot}$ (top panel) and 1 M$_{\odot}$ (bottom panel). The color
shaded regions illustrate the ($a_1$, $a_2$) pairs that can lead to nodal librations of the outer test particle in the different scenarios of work. In all cases, the
system assumes $e_1 = e_2 =$ 0.5.
  }
\label{fig:fig.a2a1}
\end{figure}

On the one hand, a simple analysis of the Eq.~\ref{eq:a2} shows that the upper limit of the semimajor axis $a_{2,\textrm{lim}}$ is an increasing function of $a_1$. By assuming values of $e_1 = e_2 =$ 0.5, the green and red curves illustrated in Fig.~\ref{fig:fig.a2a1} represent $a_{2,\textrm{lim}}$ as a function of $a_1$ for $m_1 =$ 1 M$_{\oplus}$ and $m_1 =$ 1 M$_{\textrm{Jup}}$, respectively, in systems around a central star with $m_{\text{s}} =$ 0.1 M$_{\odot}$ (top panel) and $m_{\text{s}} =$ 1 M$_{\odot}$ (bottom panel). The corresponding color shaded regions in each panel of such a figure show the ($a_1$, $a_2$) pairs that lead to nodal librations of the outer test particle. In the scenarios under consideration around a solar-mass star, an Earth-mass planet allows the existence of Kuiper belt-like reservoirs with nodal librations beyond 30 au for $a_1$ greater than 4.3 au, while a Jupiter-mass planet only requires values of $a_1$ greater than 1.2 au. For a central star of 0.1 M$_{\odot}$ in our systems of work, Kuiper belt analogues can be produced by an Earth- or a Jupiter-mass planet with $a_1$ greater than 1.55 au or 0.43 au, respectively. 

On the other hand, Eq.~\ref{eq:a2} indicates that the dependence of $a_{2,\textrm{lim}}$ on $e_1$ is somewhat more complex. To analyze this, we calculate the derivative of Eq.~\ref{eq:a2} respect to $e_1$, which is given by 

\begin{eqnarray}
    \frac{\textrm{d}a_{2,\textrm{lim}}}{\textrm{d}e_1} = \frac{4}{7}  \left(-\frac{2 a^{9/2}_{1}}{A (1 - e^{2}_{2})^{2}} \right)^{2/7} \frac{ e_{1} (3 - 8 e^{2}_{1})}{(1 + 3 e^{2}_{1} - 4 e^{4}_{1})^{5/7}}.
    \label{eq:deri_a2_e1}
\end{eqnarray}

\noindent{From this}, it is evident that this derivative vanishes for $e_1 = \sqrt{3/8}$, where Eq.~\ref{eq:a2} reaches a maximum value given by 

\begin{eqnarray}
    a^{\textrm{max}}_{2,\textrm{lim}} = 
    {\left(-\frac{25}{8} \frac{a^{9/2}_{1}}{A (1 - e^{2}_{2})^{2}} \right).}^{2/7}
    \label{eq:a2_max_e1}
\end{eqnarray}

\begin{figure}
  \centering
  \includegraphics[angle=0, width=0.5\textwidth]{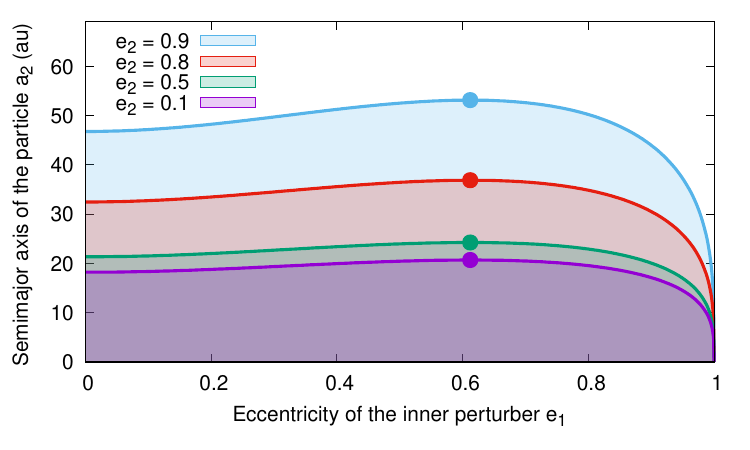}
   \caption{
Values of $a_{2,\text{lim}}$ as a function of the inner perturber's eccentricity $e_1$ for $e_2 =$ 0.1 (violet curve), 0.5 (green curve), 0.8 (red curve), and 0.9 (sky blue curve). The color circles represent the maximum value of $a_{2,\text{lim}}$, which is associated with $e_1 = \sqrt{3/8}$. The color shaded regions illustrate the ($e_1$, $a_2$) pairs that can lead to nodal librations of the outer test particle in the different scenarios of work. In all cases, the system assumes $m_{\text{s}} =$ 1 M$_{\odot}$, $m_1 =$ 1 M$_{\text{Jup}}$, and $a_1 =$ 1 au.
}
\label{fig:fig.a2_e1.eps}
\end{figure}

\noindent{By assuming a system composed of a solar-mass star and a Jupiter-mass planet with $a_1 =$ 1~au, the violet, green, red, and sky blue curves of Fig.~\ref{fig:fig.a2_e1.eps}} illustrate the values of $a_{2,\textrm{lim}}$ as a function of $e_1$ for $e_2 =$ 0.1, 0.5, 0.8, and 0.9, respectively. The corresponding color shaded regions show the ($e_1$, $a_2$) pairs that lead to nodal librations of the outer test particle. The color circles represented over each curve illustrate the maximum value of $a_{2,\textrm{lim}}$ given by Eq.~\ref{eq:a2_max_e1}. For fixed values of $m_{\text{s}}$, $m_1$, $a_1$, and $e_2$, Fig.~\ref{fig:fig.a2_e1.eps} shows that there is no strong dependency of $a_{2,\textrm{lim}}$ on the inner perturber eccentricity, except for extremely large values of $e_1$.

%===================================
\section{Numerical experiments}

Here, we present a numerical study aimed at analyzing the robustness of the analytical results derived in the previous section from a secular theory up to the quadrupole level of the approximation. First, we present the numerical code used for the development of our study. Then, we show a set of numerical experiments with and without GR effects, and a comparative analysis with the analytical results is carried out. Finally, we present applications of our study to potential outer small body reservoirs associated with real planetary systems of the observed sample.

\subsection{Numerical code}

The N-body code used for the development of the numerical experiments is that known as MERCURY \citep{Chambers1999}. Particularly, we adopt the Bulirsch-Stoer integrator with an accuracy parameter of 10$^{-12}$. As the original version of the MERCURY code only includes purely gravitational forces, we incorporate relativistic corrections adopting the first order post-Newtonian approximation proposed by \cite{Anderson1975}, which is given by

\begin{eqnarray}
\Delta {\ddot{\mathbfit{r}}} = \frac{k^2 m_{\text{s}}} {c^2 r^3} \Bigg \{ \left( \frac{4 k^2 m_{\text{s}}}{r} - {\mathbfit{v}} \cdot {\mathbfit{v}} \right) {\mathbfit{r}} + 4\left( {\mathbfit{r}} \cdot {\mathbfit{v}} \right) {\mathbfit{v}} \Bigg \},
\label{eq:Anderson}
\end{eqnarray}
where $\mathbfit{r}$ and $\mathbfit{v}$ are the astrocentric position and velocity vectors, respectively, and $r =\mid\mathbfit{r}\mid$. 

For a proper comparison with the analytical theory, the numerical results must always be referenced to barycentre and invariant plane of the system, whose x-axis is in the direction of the pericentre of the inner perturber. In absence of GR, the x-axis is fixed in space, while, when the GR effects are included in the model, we must work in a rotating frame due to the precession of the argument of pericentre of  the inner perturber.

\subsection{Comparison between N-body experiments and analytical results}

Here, we adopt a particular scenario of work with the aim of developing a comparative analysis between the analytical criteria derived in the framework of the secular quadrupolar elliptical restricted three-body problem and results of N-body numerical experiments.  

\begin{figure}
  \centering
  \includegraphics[angle=0, width=0.48\textwidth]{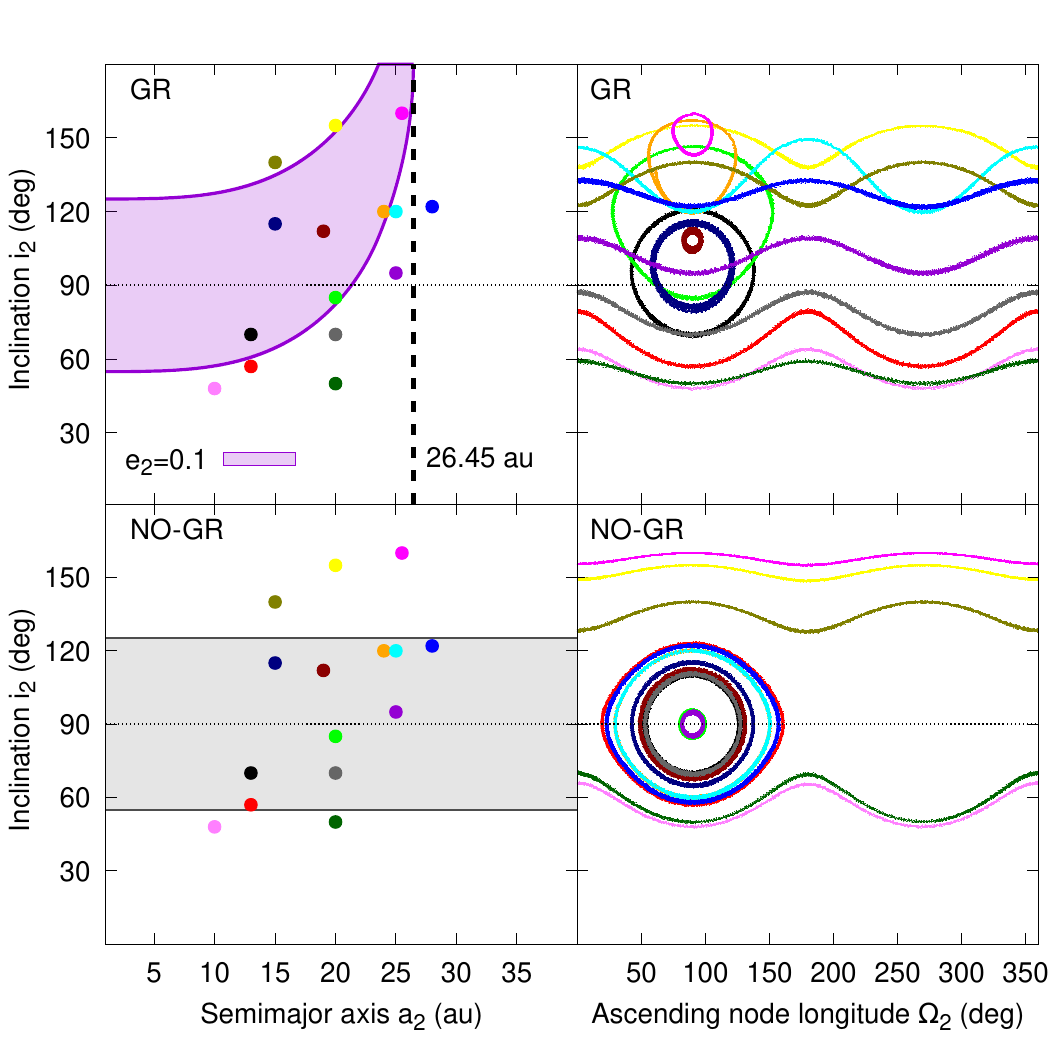}
   \caption{
Left panels: extreme inclinations and nodal libration region of an outer test particle with (top) and without (bottom) GR effects. The vertical dashed black line shows the value of $a_{2,\text{lim}}$. The color circles illustrate the initial values of $i_2$ at $\Omega_2 = 90^{\circ}$ and $a_2$ of the particles adopted for our numerical experiments. Right panels: Evolutionary trajectories in a ($\Omega_2$, $i_2$) plane of the particles, obtained from our N-body simulations with (top) and without (bottom) GR effects over 100 Myr. The system assumes $m_{\text{s}} =$ 1 M$_{\odot}$, $m_1 =$ 3 M$_{\text{Jup}}$, $a_1 =$ 1 au, $e_1 =$ 0.3 and $e_2 =$ 0.1.
   }
\label{fig:multiplot_ia_inodo_e201}
\end{figure}

\begin{figure}
  \centering
  \includegraphics[angle=0, width=0.48
  \textwidth]{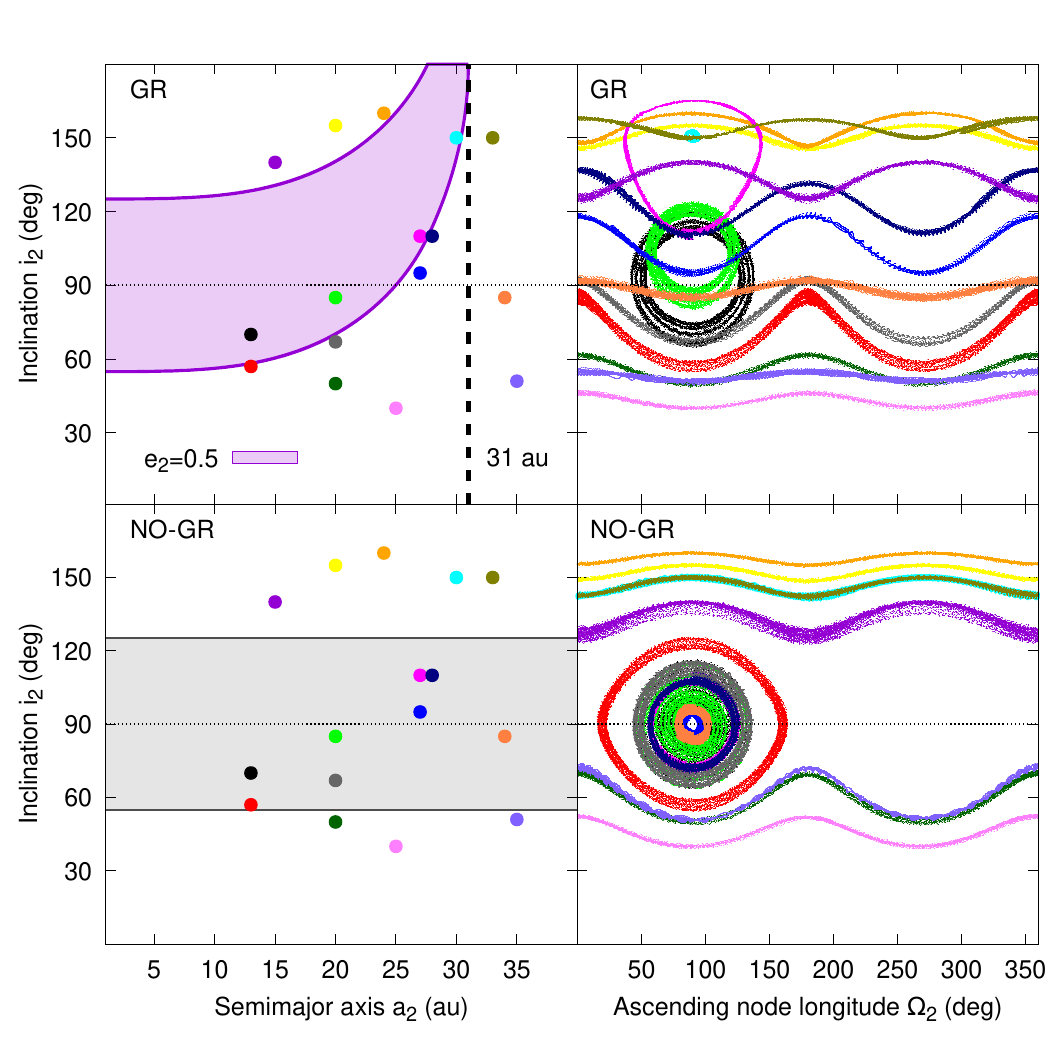}
   \caption{
   The physical and orbital parameters of the massive bodies and the references associated with regions, curves, lines, and circles illustrated in these panels are described in the caption of Fig.~\ref{fig:multiplot_ia_inodo_e201}, but considering massless particles with $e_2 =$ 0.5 in this case. 
}
\label{fig:multiplot_ia_inodo_e205}
\end{figure}

Figures~\ref{fig:multiplot_ia_inodo_e201} and \ref{fig:multiplot_ia_inodo_e205} illustrate the numerical integration of a set of test particles associated with a system composed of a solar-mass star, an inner perturber with $m_1 =$ 3~M$_{\text{Jup}}$, $a_1 =$ 1~au, and $e_1 =$ 0.3, for $e_2 =$ 0.1 and 0.5, respectively. One the one hand, the top left panels of both figures show the violet shaded nodal libration region and the variable extreme inclinations by violet curves in a ($a_2$, $i_2$) plane with GR effects. At the same way, the bottom left panels of both figures illustrate the gray shaded nodal libration region and the fixed extreme inclinations by gray horizontal lines in a ($a_2$, $i_2$) plane without GR effects. Moreover, such panels illustrate the initial values of $a_2$ and $i_2$ adopted for the test particles by color circles. It is important to remark that such values of $i_2$ are associated with $\Omega = 90^{\circ}$. On the other hand, the right panels of Figs.~\ref{fig:multiplot_ia_inodo_e201} and \ref{fig:multiplot_ia_inodo_e205} show the dynamical evolution of the test particles during 100 Myr in the ($\Omega_2$, $i_2$) plane with (top) and without (bottom) GR effects. 

Our study shows very good agreement between the analytical criteria described in Sect. 3 and the results obtained from N-body experiments for all simulated test particles with and without GR effects.

In absence of GR, the evolution of the test particles illustrated in the bottom right panels of Figs.~\ref{fig:multiplot_ia_inodo_e201} and \ref{fig:multiplot_ia_inodo_e205} are consistent with the analytical results derived in the pioneering work of \citet{Ziglin1975}. In fact, on the one hand, test particles with initial conditions inside the nodal libration region experience oscillations of $\Omega_2$ coupled with ﬂips of the orbital plane from prograde to retrograde values of $i_2$ and back again along the evolution. For these particles, it is worth remarking that the oscillations of the orbital plane are symmetric around $i_2 =$ 90$^{\circ}$, where $\Omega_2$ adopts extreme values. On the other hand, all test particles with initial conditions outside the nodal libration region show circulations of $\Omega_2$ with purely prograde or retrograde orbits. As the reader can see, these test particles have initial values of $i_2$ at $\Omega_2 = 90^{\circ}$ smaller or higher than the minimum or maximum extreme inclinations, respectively, which are calculated from Eq.~\ref{eq:incli_ziglin}. 
Finally, it is important to remark that there is no constraint to the semimajor axis of outer test particles that experience oscillations of $\Omega_2$ in absence of GR, since the nodal libration region only depends on the inner perturber eccentricity's $e_1$ \citep{Ziglin1975}.

When the GR effects are included in the model, the evolution of the test particles showed in the top right panels of Figs.~\ref{fig:multiplot_ia_inodo_e201} and \ref{fig:multiplot_ia_inodo_e205} are in a very good agreement with the analytical results derived by \citet{Zanardi2018} as well as with those obtained in the present work. On the one hand, the test particles with initial conditions inside the nodal libration region show oscillations of $\Omega_2$ along their evolution. As proposed in the pioneering work developed by \citet{Zanardi2018}, we observe two different regimes of nodal libration in our simulated test particles, which depend on the evolution of $i_2$. First, oscillations of $\Omega_2$ correlated with orbital flips from prograde to retrograde values of $i_2$ and back again. In this scenario of work, all simulated test particles show asymmetric librations of $i_2$ around 90$^{\circ}$. Second, oscillations of $\Omega_2$ associated with purely retrograde orbits throughout the entire evolution. As shown by Eq.~\ref{eq:iestre}, the extreme of $\Omega_2$ are always associated with retrograde values of $i_2$ due to GR effects, which allows to understand the peculiar characteristic of the simulated test particles that experience nodal librations in the two different regimes previously mentioned. On the other hand, all test particles of our numerical experiments with initial conditions outside the nodal libration region show circulations of $\Omega_2$. As the reader can see, these test particles have an initial semimajor axis greater than $a_{2,\text{lim}}$ given by Eq.~\ref{eq:a2}, or an initial semimajor axis less than such $a_{2,\text{lim}}$ and initial values of $i_2$ at $\Omega_2 = 90^{\circ}$ smaller or higher than the minimum or maximum extreme inclinations, respectively, which are calculated from Eqs.~\ref{eq:cuadratica} and \ref{eq:cuadratica_nueva}. Test particles with nodal circulations whose initial semimajor axis is greater than $a_{2,\text{lim}}$ represent very good examples that illustrate how the GR effects impose constraints to the semimajor axis of outer test particles with nodal librations.

%----------

\begin{figure*}
  \centering
  \includegraphics[angle=0, width=\textwidth]{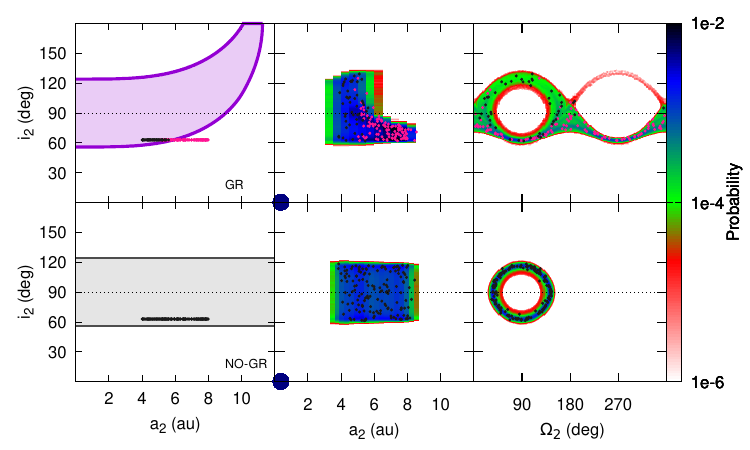}
   \caption{N-body experiments for the system HIP 79431. The left panels illustrate the initial values of $i_2$ at $\Omega_2 = 90^{\circ}$ and $a_2$ of the test particles adopted for the simulations. Initial conditions that lead to librations and circulations of $\Omega_2$ are represented by dark gray and dark pink circles, respectively. The extreme inclinations and the nodal libration region for $e_2 =$ 0.5 are also illustrated. The middle and right panels show evolutionary maps of occupation in the ($a_2$, $i_2$) and ($\Omega_2$, $i_2$) planes, respectively, where the color regions illustrate zones where the test particles can be found with different probability levels. Moreover, orbital parameters at 100 Myr of particles with initial conditions associated with librations and circulations of $\Omega_2$ are shown by dark gray and dark pink circles, respectively. The blue circles in the middle panels represent the planet of the system. Finally, we remark that the top and bottom panels are associated with scenarios with and without GR effects, respectively. 
}
\label{fig:mapa_hip79431}
\end{figure*}  

\subsection{Dynamical properties of potential outer reservoirs around HIP 79431 and GJ 514}
 
Here, we carry out N-body simulations with the aim of studying the global dynamical structure of hypothetical outer small body populations associated with two different real systems hosting an eccentric inner planet around a single stellar component. The systems selected from the observed sample are HIP 79431 and GJ 514, which cover a wide range of values of the mass $m_1$ of the planetary companion. In each of these systems, we model the hypothetical reservoirs by assuming massless particles with initial eccentricities $e_2 <$ 0.5 and a ratio of semimajor axes $\alpha = \frac{a_1}{a_2} \lesssim 0.1$ in order to analyze secular interactions up to quadrupole order. The initial values of the inclination $i_2$ of the test particles are particularly chosen for each system of study. We assume that the orbital characteristics of the hypothetical reservoirs could have originated from planetary scattering events during the evolution of the systems under consideration.
% ----------------------------------

\subsubsection{The case of HIP 79431}

HIP 79431 is a system composed of a central star of 0.49~M$_\odot$, which is orbited by a giant planet with a minimum mass of 2.1~M$_\text{Jup}$, semimajor axis of 0.36~au, and eccentricity of 0.29 \citep{Apps2010}. To model the hypothetical reservoir of small bodies, we assume a set of massless particles with initial semimajor axes $a_2$ between 4~au and 8~au, and an initial inclination $i_2$ at $\Omega_2 = 90^{\circ}$ of 63$^{\circ}$.

The left panels of Fig.~\ref{fig:mapa_hip79431} represent the initial values of $a_2$ and $i_2$ at $\Omega_2 = 90^{\circ}$ of the test particles adopted for our numerical experiments. The dark gray and dark pink circles illustrate particles with initial conditions that lead to librations and circulations of $\Omega_2$, respectively. Moreover, the extreme inclinations and the nodal libration region with (top left panel) and without (bottom left panel) GR for $e_2 =$ 0.5 are included as reference. On the one hand, all test particles of the N-body simulations without GR have initial conditions associated with nodal libration trajectories. On the other hand, a set of test particles of the N-body experiments with GR with 4 au $\leq a_2 \lesssim$ 5.6 au have initial conditions consistent with librations of $\Omega_2$, while another set of test particles with 5.1 au $\lesssim a_2 \leq$ 8 au show initial conditions associated with circulations of $\Omega_2$.    

Figure~\ref{fig:mapa_hip79431} also illustrates the dynamical evolution along 100 Myr of the test particles of the outer reservoir in the planes ($a_2$, $i_2$) (middle panels) and ($\Omega_2$, $i_2$) (right panels) with (top panels) and without (bottom panels) GR effects. Such evolution is shown by occupation maps, which represent the normalized time in which the test particles reside in the different zones of the aforementioned planes during the entire integration time. The color code indicates the time spent in each region, being blue for the most visited region and red for the least visited one. Moreover, the blue circles illustrated in the middle panels represent the planet of the system. Finally, in the middle and right panels, the dark gray and dark pink circles illustrate the orbital parameters at 100 Myr of test particles with initial conditions associated with librations and circulations of $\Omega_2$, respectively.

The middle and right panels of Fig.~\ref{fig:mapa_hip79431} show that the evolution of the test particles with and without GR effects is consistent with their initial conditions. On the one hand, in absence of GR, all test particles experience nodal librations with orbital flips over 100 Myr of evolution. On the one hand, when the GR is considered, an inner test particle population with nodal librations and orbital flips coexists with an outer one with nodal circulations and purely prograde orbits at 100 Myr. The transition between these two populations with different nodal evolution regimes occurs at semimajor axes between 5.1 au and 5.6 au, where test particles with initial conditions consistent with librations and circulations of $\Omega_2$ coexist throughout the entire evolution.
% ----------------------------------

\subsubsection{The case of GJ 514}

\begin{figure*}
  \centering
  \includegraphics[angle=0, width=\textwidth]{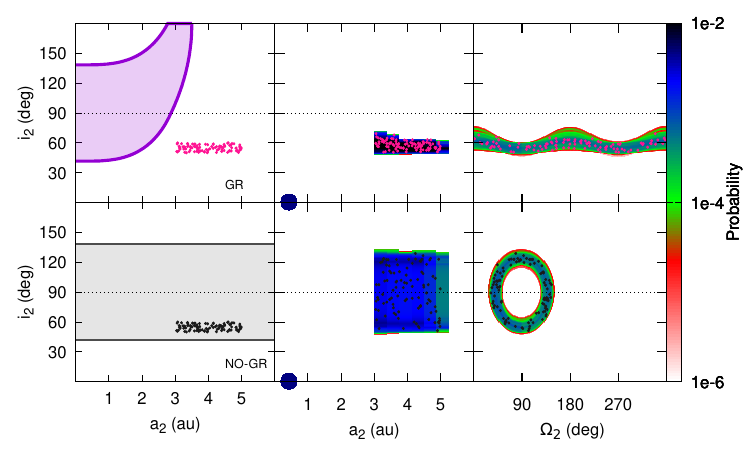}
   \caption
   {N-body experiments for the system GJ 514. The references associated with regions, curves, circles, and occupation maps illustrated in these panels are described in the caption of Fig.~\ref{fig:mapa_hip79431}.
   }
\label{fig:mapa_gj514}
\end{figure*}

GJ 514 is a system composed of a central star of 0.51 M$_\odot$, and a planetary companion residing in the habitable zone with a minimum mass of 5.2 M$_{\oplus}$, a semimajor axis of 0.422~au, and an eccentricity of 0.45 \citep{Damasso2022}. For this system, we consider a population of massless particles with initial semimajor axes $a_2$ between 3~au and 5~au, and initial values of the inclination $i_2$ at $\Omega_2 = 90^{\circ}$ ranging from 50$^\circ$ to 60$^\circ$.

The left panels of Fig.~\ref{fig:mapa_gj514} show the initial values of $a_2$ and $i_2$ at $\Omega_2 = 90^{\circ}$ of the test particles of our numerical experiments. The dark gray and dark pink circles represent particles with initial conditions associated with librations and circulations of $\Omega_2$, respectively. As in Fig.~\ref{fig:mapa_hip79431}, the extreme inclinations and the nodal libration region with (top left panel) and without (bottom left panel) GR for $e_2 =$ 0.5 are represented as reference. In this particular case, all test particles of the N-body experiments with and without GR effects have initial conditions associated with circulations and librations of $\Omega_2$, respectively.

The middle and right panels of Fig.~\ref{fig:mapa_gj514} illustrate occupation maps, which describe the dynamical evolution along 100 Myr of the test particles in the ($a_2$, $i_2$) and ($\Omega_2$, $i_2$) planes, respectively, with (top panels) and without (bottom panels) GR effects. In addition, the blue circles in the middle panels represents the planet of the system, while the circles illustrated in the middle and right panels show the orbital parameters of the test particles at 100 Myr, where dark gray and dark pink are assigned to initial conditions of librations and circulations of $\Omega_2$, respectively. 

Our results show an excellent agreement between the evolution of the test particles over 100 Myr and their initial conditions in all N-body experiments. In fact, in absence of GR, all particles evolve on nodal libration trajectories with orbital flips throughout the entire simulation. On the contrary, all particles of the numerical experiments with GR experience nodal circulations with prograde inclinations, which keep values close to the initial one during 100 Myr.  
          
\vspace{0.5cm}

A simple analysis of the systems HIP 79431 and GJ 514 show how the GR effects can significantly modify the global dynamical structure of hypothetical outer reservoirs, which depends on the initial conditions of the test particles adopted for their modeling. From these results, we want to emphasize the need to incorporate GR effects in models of structure and evolution of debris disks aimed at understanding the global dynamics of real exoplanetary systems.

%======================
\section{Discussion and conclusions}

In this research, we present a detailed study about how the GR effects modify the dynamical properties of an outer test particle in the framework of the elliptical restricted three-body problem. We provide improved analytical expressions at the quadrupole level of the secular approximation that allow to determine the nodal libration region of an outer test particle with GR, which depends on the physical and orbital parameters of the bodies of the system. 

One of the main results of our study is the existence of an upper limit of the semimajor axis $a_{2,\text{lim}}$ below which an outer test particle can experience nodal librations. According to our analysis, the value of $a_{2,\text{lim}}$ depends on the mass of the central star $m_{\text{s}}$, the mass $m_1$, the semimajor axis $a_1$, and the eccentricity $e_1$ of the inner perturber, and the eccentricity $e_2$ of the outer test particle. On the one hand, our results show that the greater $m_1$, $a_1$ and $e_2$ and the smaller $m_\text{s}$, the greater the value of $a_{2,\text{lim}}$. On the other hand, we find a particular $e_1 = \sqrt{3/8}$ where $a_{2,\text{lim}}$ reaches a maximum value. However, except for large $e_1$, $a_{2,\text{lim}}$ does not strongly depend on such an orbital parameter for fixed values of $m_\text{s}$, $m_1$, $a_1$, and $e_2$.

The results obtained from N-body experiments that include GR effects are consistent with the analytical prescriptions described in Sects. 2 and 3 of the present work. Moreover, the analysis of hypothetical outer reservoirs associated with real exoplanetary systems that host an inner planet around a single stellar component shows that the GR can play a key role in their global dynamical structure. 

We want to remark two important points in the framework of our research. First, as found by \citet{Zanardi2018}, our study shows that the GR leads to two different regimes of nodal libration for outer test particles in the elliptical restricted three-body problem, which depend on the evolution of their orbital inclination. In fact, in this scenario of work, the nodal librations can be correlated with orbital flips or purely retrograde orbits of the outer test particle. We consider that studying these nodal libration regimes as a function of the semimajor axis of the outer test particle is of priority interest. In fact, this research will allow us to complete our understanding concerning the global dynamical structure of the systems under consideration for a wide diversity of physical and orbital parameters of the bodies that compose them. Second, in addition to GR, the precession of the argument of pericentre of the inner orbit could also arise from other effects such as tides and rotational deformation \citep{Sterne1939}. The consideration of these effects in the modeling of the apsidal precession rate of the inner orbit could modify the analytical prescriptions of the present study that define the nodal libration region of outer test particles in the elliptical restricted three-body problem. The topics of interest described in this paragraph will be the focus of our research in forthcoming papers.

The results obtained in the present research strengthen the paradigm concerning the primary role that GR plays in the global dynamic structure of systems that host an inner planet. Its inclusion in dynamical models will be essential to access a correct interpretation of the structure of outer debris disks as well as a better understanding of the cold dust generation rate in those scenarios.

\section*{Acknowledgements}
We thank the anonymous referee for her/his comments and suggestions.
The authors acknowledge the partial support by Universidad Nacional de La Plata, Argentina, through PID G172. Moreover, they acknowledge the financial support by Facultad de Ciencias Astronómicas y Geofísicas de La Plata and Instituto de Astrofísica de La Plata for extensive use of their computing facilities. In particular, M.Z. thanks the support of the Agencia Nacional de Promoción de la Investigación, el Desarrollo Tecnológico y la Innvovación, Argentina, through the PICT 2019-2312. Finally, M.Z. and G.C.dE wish to dedicate the present paper to Lourdes de El\'ia, main source of love and inspiration of each day.

%%%%%%%%%%%%%%%%%%%%%%%%%%%%%%%%%%%%%%%%%%%%%%%%%%
\section*{Data Availability}
%The inclusion of a Data Availability Statement is a requirement for articles published in MNRAS. Data Availability Statements provide a standardised format for readers to understand the availability of data underlying the research results described in the article. The statement may refer to original data generated in the course of the study or to third-party data analysed in the article. The statement should describe and provide means of access, where possible, by linking to the data or providing the required accession numbers for the relevant databases or DOIs.

The numerical experiments shown in this reasearch will be shared with interested readers upon request to the first author of the article.

%%%%%%%%%%%%%%%%%%%% REFERENCES %%%%%%%%%
\bibliographystyle{mnras}
\bibliography{Zanardi2023} % if your bibtex file is called example.bib
%\end{thebibliography}

%%%%%%%%%%%%%%%%% APPENDICES %%%%%%%%
%\appendix
%%%%%%%%%%%%%%%%%%%%%%%%%%%%%%%%%%%%%
% Don't change these lines
\bsp	% typesetting comment
\label{lastpage}
\end{document}